\documentclass[prd,eqsecnum,preprint,tightenlines,nofootinbib,showpacs]{revtex4}
\usepackage{graphicx,amsmath,latexsym}

\def\bea{\begin{eqnarray}}
\def\eea{\end{eqnarray}}
\def\be{\begin{equation}}
\def\ee{\end{equation}}
\newcommand{\ub}[1]{\underline{#1}}
\def\del{\partial}

\def\ths{\thinspace}
\def\ha{\frac{1}{2}}
\def\senk#1{\vec{#1}_\perp}

\begin{document}

\title{Restoration of the chiral limit \\
in Pauli--Villars-regulated light-front QED}

\author{Sofia S. Chabysheva}
\affiliation{Department of Physics \\
Southern Methodist University \\
Dallas, Texas 75275 \\
and \\
Department of Physics \\
University of Minnesota-Duluth \\
Duluth, Minnesota 55812}

\author{John R. Hiller}
\affiliation{Department of Physics \\
University of Minnesota-Duluth \\
Duluth, Minnesota 55812}

\date{\today}

\begin{abstract}
The dressed-electron eigenstate of Feynman-gauge QED is computed
in light-front quantization with a Fock-space truncation to 
include at most the one-photon/one-electron sector.  The theory
is regulated by the inclusion of three massive Pauli--Villars (PV) particles,
one PV electron, and two PV photons.  In particular,
the chiral limit is investigated, and the correct limit is found
to require two PV photons, not just one as previously thought.
The renormalization and covariance of the electron current are
also analyzed.  We find that the plus component is well behaved
and use its spin-flip matrix element to compute the electron's
anomalous moment.  The dependence of the moment on the regulator
masses is shown to be slowly varying when the second PV photon
is used to guarantee the correct chiral limit.
\end{abstract}

%
\pacs{12.38.Lg, 11.15.Tk, 11.10.Gh, 11.10.Ef
%
}

\maketitle

\section{Introduction}
\label{sec:Introduction}

Over the past several years, a method of Pauli--Villars (PV) 
regularization~\cite{PauliVillars} has been developed for
nonperturbative analysis of quantum field theories~\cite{bhm1,bhm2,%
YukawaDLCQ,ExactSolns,YukawaOneBoson,OnePhotonQED,YukawaTwoBoson}.  
It is based on the introduction of massive, negatively normed fields 
directly to the Lagrangian and the derivation of a light-front
quantized Hamiltonian~\cite{Dirac,DLCQreview}.  The Hamiltonian
is then used to construct an eigenvalue problem for the mass and
Fock-state wave functions of bound states.  The use of light-front
quantization allows a meaningful Fock-state expansion with
well-defined wave functions.

Bound-state problems in quantum field theories are notoriously
difficult.  Their nonperturbative nature complicates the
regularization and renormalization.  Of the various methods
that have been attempted, such as lattice gauge theory~\cite{lattice},
Schwinger--Dyson equations~\cite{SDE}, and light-front 
quantization~\cite{DLCQreview}, only the light-front approach 
can provide well-defined wave functions.

To regulate the nonperturbative light-front problem, the regulators
that work in perturbation theory and provide equivalence with 
Feynman perturbation theory are assumed to be sufficient.  Careful
studies of perturbative equivalence have been made by 
Paston {\em et al}~\cite{Paston}.  To renormalize, the bare parameters
of the Lagrangian are fixed via physical conditions, such as setting
certain bound-state masses equal to measured values.  This is
distinct from the sector-dependent approach~\cite{SectorDependent}
to renormalization, where the bare parameters are assigned different
values in each Fock sector.

The purpose in studying QED with such a technique is to test methods
in a gauge theory, the goal being to develop a method that works for
bound states of QCD.  There is no expectation that nonperturbative
light-front results for QED will be at all competitive with high-order
perturbative calculations~\cite{Kinoshita}.  The numerical errors
in solving the bound-state eigenvalue problem are currently of
order 1\%.  For the calculation reported here, the Fock-space
truncation is severe enough to make calculations tractable analytically,
but not enough physics is included to expect close agreement with
experiment.  The main point of the calculation
is instead that the behavior of the anomalous moment is now a slowly
varying function of the regulator masses.

Specifically, we reconsider the dressed-electron state in Feynman-gauge
QED truncated at the one-photon/one-electron Fock state.
An earlier analysis~\cite{OnePhotonQED} was sufficient in a
particular limit of Pauli--Villars masses.  There, one PV electron
and one PV photon were added to the Lagrangian to regulate the
theory.  The resulting bound-state problem was solved analytically,
and the anomalous moment was calculated in the limit that the
PV electron mass is taken to infinity.  

When the PV electron
mass is not infinite, however, the analysis breaks down, due
to a violation of chiral symmetry in the massless electron limit. 
This violation was not recognized in the earlier work on this 
particular PV regularization~\cite{OnePhotonQED} but is quite
consistent with what has been found in different PV regularizations
of QED and Yukawa theory~\cite{ChiralSymYukawa}. 
We restore this symmetry by adding a
second PV photon, with its coupling strength and mass related
by a simple condition.  We also verify that the electron one-loop
self-energy is consistent with the Feynman result at the same
order and show that the vertex and wave function renormalization 
constants, $Z_1$ and $Z_2$, are equal.  However, the equality
of $Z_1$ and $Z_2$ is effective only for the plus component
of the current; our truncation destroys the covariance of
the current, and only the plus component can be used.

With the second PV photon included, we are able to do a
calculation of the electron's anomalous moment at finite
PV electron mass.  The moment is computed from the spin-flip
matrix element of the (unrenormalized) plus component of the 
current~\cite{BrodskyDrell}.  We show that without the
second PV photon, the anomalous moment has a strong 
dependence on the PV masses, and we identify the mechanism
whereby restoration of the correct chiral limit removes the
strong mass dependence.

The alternative, sector-dependent renormalization 
scheme~\cite{SectorDependent} 
has been employed in studies of light-front
QED by Hiller and Brodsky~\cite{hb} and more recently by 
Karmanov {\em et al}.~\cite{Karmanov:2006zf,Bernard}.
Unfortunately, the most recent form of the 
sector-dependent approach~\cite{Karmanov:2006zf} 
leads to difficulties with the interpretation of the wave functions.
The amplitude for the bare-electron state is of the form
$\sqrt{1-\alpha J}$, where $J$ is a positive integral that is infinite
when the regulators are removed.  The probability of the
one-photon/one-electron sector is $\alpha J$, and thus the
norm is 1.  For our case, the bare amplitude is $1/\sqrt{1+\alpha J}$
and the probability of the higher sector is $\alpha J/(1+\alpha J)$,
again with a norm of 1.  However, only in our case are there
well-defined probabilities between zero and 1 for each
Fock sector and for any value of $J$.  What is more, if one calculates
something directly from the wave functions of Karmanov {\em et al}.,
infinities are encountered before taking the regulator masses to infinity.  
For example, the expectation value of
the number of photons in the dressed-electron state is infinite
for finite PV masses, even in the one-photon truncation.  An
analogous calculation in QCD, such as a quark distribution function,
would also yield infinity.  In their method, useful information
can be extracted from the wave functions only by embedding the
eigenstate in a larger process and using an external probe
combined with a separate renormalization of this external
coupling, a process not so different from the effort required
in lattice QCD, where wave functions are also not well defined.

Our calculations are done in terms of light-cone coordinates~\cite{Dirac},
which are defined by
\begin{equation}
x^\pm \equiv x^0 \pm x^3, \;\; \senk{x}\equiv (x^1,x^2).
\end{equation} 
The covariant four-vector is written $x^\mu = (x^+,x^-,\senk{x})$.
This corresponds to a spacetime metric of
\begin{equation}
g^{\mu\nu}=\left(\begin{array}{llll} 0&2&0&0\\
		   2&0&0&0\\
		   0&0&-1&0\\
		   0&0&0&-1 \end{array} \right) .
\end{equation}
Dot products are then given by
\begin{equation}
x\cdot y=g_{\mu\nu}x^\mu y^\nu = \ha(x^+y^-+x^-y^+)-\senk{x}\cdot\senk{y}.
\end{equation} 
For light-cone three-vectors we use the underscore notation
\begin{equation}
\ub{x} \equiv (x^-,\senk{x}) .
\end{equation} 
For momentum, the conjugate to $x^-$ is $p^+$, and, therefore, we use
\begin{equation}
\ub{p} \equiv (p^+,\senk{p})
\end{equation} 
as the light-cone three-momentum.
The dot product of momentum and position three-vectors is
\begin{equation}
\ub{p} \cdot\ub{x} \equiv\ha p^+x^- - \senk{p}\cdot\senk{x} .
\end{equation} 
The derivatives are
\begin{equation}
\del_+ \equiv \frac{\del}{\del x^+}\ths,\qquad
\del_- \equiv \frac{\del}{\del x^-}\ths,\qquad
\del_{i} \equiv \frac{\del}{\del x^i} .
\end{equation} 

The time variable is taken to be $x^+$, and time evolution of a system
is then determined by ${\cal P}^-$, the operator associated with the
momentum component conjugate to $x^+$.  Stationary
states are obtained as eigenstates of ${\cal P}^-$.
As has been customary, we express the eigenvalue
problem in terms of a light-cone 
Hamiltonian~\cite{PauliBrodsky,DLCQreview}
$H_{\rm LC}={\cal P}^+{\cal P}^-$ as
\begin{equation} \label{eq:EigenProb}
H_{\rm LC}|P\rangle=(M^2+P_\perp^2)|P\rangle,\;\;
\underline{\cal P}|P\rangle=\underline{P}|P\rangle,
\end{equation}
where $M$ is the mass of the state, and ${\cal P}^+$ and
$\vec{\cal P}_\perp$ are light-cone momentum operators.
Without loss of generality, we will 
limit the total transverse momentum $\vec{P}_\perp$ to zero.

The structure of the remainder of the paper is as follows.
In Sec.~\ref{sec:FeynmanGauge} we summarize the Feynman-gauge
formulation of light-front QED, including two PV photons and
one PV electron as regulators, and construct the Hamiltonian
that defines the bound-state problem.  We then give in 
Sec.~\ref{sec:Truncation} an update of the known analytic
solution of the one-photon truncation~\cite{OnePhotonQED} 
to include the additional PV
photon.  The analysis of the electron self-energy, in 
particular the chiral limit, and of the renormalization of
the current are discussed in Sec.~\ref{sec:Regularization}.
These developments are applied to the calculation of
the anomalous moment in Sec.~\ref{sec:AnomalousMoment},
followed by a brief summary in Sec.~\ref{sec:Summary}.
Appendices contain a discussion of the gauge condition
and a proof of a useful identity for 
terms in the self-energy.

\section{Light-Front QED in Feynman Gauge}
\label{sec:FeynmanGauge}

The Feynman-gauge QED Lagrangian, regulated by two PV photons and 
one PV electron, is
\bea \label{eq:Lagrangian}
{\cal L} &= & \sum_{i=0}^2 (-1)^i \left[-\frac14 F_i^{\mu \nu} F_{i,\mu \nu} 
         +\frac12 \mu_i^2 A_i^\mu A_{i\mu} 
         -\frac{1}{2} \left(\partial^\mu A_{i\mu}\right)^2\right] \\
  && + \sum_{i=0}^1 (-1)^i \bar{\psi_i} (i \gamma^\mu \partial_\mu - m_i) \psi_i 
  - e \bar{\psi}\gamma^\mu \psi A_\mu ,  \nonumber
\eea
where
\begin{equation} \label{eq:NullFields}
  A_\mu  = \sum_{i=0}^2 \sqrt{\xi_i}A_{i\mu}, \;\;
  \psi =  \sum_{i=0}^1 \psi_i, \;\;
  F_{i\mu \nu} = \partial_\mu A_{i\nu}-\partial_\nu A_{i\mu} .
\end{equation}
The subscript $i=0$ denotes a physical field and $i=1$ or 2 a PV
field.  Fields with odd index $i$ are chosen to be negatively
normed.  The mass of the physical photon, $\mu_0$, is set to zero.

The constants $\xi_i$ are introduced to adjust the couplings
$\sqrt{\xi_i}e$ of the different photon flavors, in order to
arrange the cancellations that are at the heart of PV
regularization.  The $\xi_i$ must satisfy constraints.  One is
simply that $\xi_0=1$, so that $e$ is the coupling of the physical
electron to the physical photon.  Another is to guarantee that
summing over photon flavors, in an internal line of a 
Feynman graph, cancels the
leading divergence associated with integration over the 
momentum of that line.  Since the $i$th flavor has norm $(-1)^i$
and couples to a charge $\sqrt{\xi_i}e$ at each end, the constraint
is
\be \label{eq:nullconstraint}
\sum_{i=0}^2(-1)^i\xi_i=0 .
\ee
This also guarantees that $A^\mu$ in (\ref{eq:NullFields}) is a zero-norm field.  
A third constraint will be imposed in Sec.~\ref{sec:Regularization},
to obtain the correct chiral limit.

The dynamical fields are
\bea
\psi_{i+}&=&\frac{1}{\sqrt{16\pi^3}}\sum_s\int d\ub{k} \chi_s
  \left[b_{is}(\ub{k})e^{-i\ub{k}\cdot\ub{x}}
        +d_{i,-s}^\dagger(\ub{k})e^{i\ub{k}\cdot\ub{x}}\right]\,, \\
A_{i\mu}&=&\frac{1}{\sqrt{16\pi^3}}\int \frac{d\ub{k}}{\sqrt{k^+}}
  \left[a_{i\mu}(\ub{k})e^{-i\ub{k}\cdot\ub{x}}
        +a_{i\mu}^\dagger(\ub{k})e^{i\ub{k}\cdot\ub{x}}\right]\,,
\eea
with~\cite{LepageBrodsky} $\chi_s$ an eigenspinor of $\Lambda_+\equiv\gamma^0\gamma^+/2$.
The creation and annihilation operators satisfy (anti)commutation
relations
\bea
\{b_{is}(\ub{k}),b_{i's'}^\dagger(\ub{k}'\}
   &=&(-1)^i\delta_{ii'}\delta_{ss'}\delta(\ub{k}-\ub{k}'), \\
\{d_{is}(\ub{k}),d_{i's'}^\dagger(\ub{k}'\}
   &=&(-1)^i\delta_{ii'}\delta_{ss'}\delta(\ub{k}-\ub{k}'), \\
{[}a_{i\mu}(\ub{k}),a_{i'\nu}^\dagger(\ub{k}']
   &=&(-1)^i\delta_{ii'}\epsilon^\mu\delta_{\mu\nu}\delta(\ub{k}-\ub{k}').
\eea
Here $\epsilon^\mu = (-1,1,1,1)$ is the metric signature for the
photon field components in Gupta--Bleuler quantization~\cite{GuptaBleuler,GaugeCondition}.
For the zero-norm photon field $A_\mu$, we have $a_\mu=\sum_i\sqrt{\xi_i}a_{i\mu}$
and the commutator
\be
{[}a_\mu(\ub{k}),a_\nu^\dagger(\ub{k}')]
   =\left[\sum_i (-1)^i\xi_i\right]\epsilon^\mu\delta_{\mu\nu}\delta(\ub{k}-\ub{k}')=0.
\ee
The implementation of the gauge condition $\partial^\mu A_{i\mu}=0$ is discussed
in Appendix~\ref{sec:GaugeCondition}.

An important consequence of the regularization method
is that one is not limited to light-cone gauge.  
The coupling of the two zero-norm fields $A^\mu$ and $\psi$
as the interaction term
reduces the fermionic constraint equation
to a solvable equation without forcing the gauge field $A_-=A^+$ to zero.
The nondynamical components of the fermion fields satisfy the
constraints ($i=0,1$)
\bea \label{eq:FermionConstraint}
i(-1)^i\partial_-\psi_{i-}+e A_-\sum_j\psi_{j-} 
  &=&(i\gamma^0\gamma^\perp)
     \left[(-1)^i\partial_\perp \psi_{i+}-ie A_\perp\sum_j\psi_{j+}\right] 
     \nonumber  \\
   &&  -(-1)^i m_i \gamma^0\psi_{i+} . 
\eea
It would appear that a nontrivial inversion of the covariant
derivative is needed to solve these constraints, except when
light-cone gauge ($A^+=0$) is used.  However, if we subtract
(\ref{eq:FermionConstraint}) for $i=1$ from (\ref{eq:FermionConstraint})
for $i=0$, the terms containing the gauge field cancel, and the constraint 
reduces to
\be
i\partial_-(\psi_{0-}+\psi_{1-})
  =(i\gamma^0\gamma^\perp)
     \partial_\perp (\psi_{0+}+\psi_{1+})
      - \gamma^0(m_0\psi_{0+}+m_1\psi_{1+}) .
\ee
Thus, the nondynamical part of the null combination $\psi_0+\psi_1$
that couples to $A^+$ satisfies the same constraint as does the
free fermion field.  This constraint is then solved explicitly, and the
nondynamical fermion fields are eliminated from the Lagrangian.
The full Fermi field can then be written as
\be
\psi_i=\frac{1}{\sqrt{16\pi^3}}\sum_s\int \frac{d\ub{k}}{\sqrt{k^+}} 
  \left[b_{is}(\ub{k})e^{-i\ub{k}\cdot\ub{x}}u_{is}(\ub{k})
        +d_{i,-s}^\dagger(\ub{k})e^{i\ub{k}\cdot\ub{x}}v_{is}(\ub{k})
        \right] ,
\ee
and the light-cone Hamiltonian ${\cal P}^-$ can be constructed directly 
from the above Lagrangian.

Another important consequence of the regularization
is the absence of instantaneous fermion
contributions.  The contributions from the instantaneous physical electron
and the instantaneous PV electron cancel, because they are of opposite
sign and are independent of the fermion mass.

The regularization scheme does have
the disadvantage of breaking gauge invariance, through the
presence of ``flavor'' changing currents where a physical
fermion can be transformed to a PV fermion or vice versa.
However, the breaking
effects disappear in the limit of large PV fermion 
mass~\cite{OnePhotonQED}, because the physical fermion
cannot make a transition to a state with infinite mass.

Without antifermion terms, the Hamiltonian is
\begin{eqnarray} \label{eq:QEDP-}
\lefteqn{{\cal P}^-=
   \sum_{i,s}\int d\ub{p}
      \frac{m_i^2+p_\perp^2}{p^+}(-1)^i
          b_{i,s}^\dagger(\ub{p}) b_{i,s}(\ub{p})} \\
   && +\sum_{l,\mu}\int d\ub{k}
          \frac{\mu_l^2+k_\perp^2}{k^+}(-1)^l\epsilon^\mu
             a_{l\mu}^\dagger(\ub{k}) a_{l\mu}(\ub{k})
          \nonumber \\
   && +\sum_{i,j,l,s,\mu}\int d\ub{p} d\ub{q}\left\{
      b_{i,s}^\dagger(\ub{p}) \left[ b_{j,s}(\ub{q})
       V^\mu_{ij,2s}(\ub{p},\ub{q})\right.\right.\nonumber \\
      &&\left.\left.\rule{0.5in}{0in}
+ b_{j,-s}(\ub{q})
      U^\mu_{ij,-2s}(\ub{p},\ub{q})\right] 
            \sqrt{\xi_l}a_{l\mu}^\dagger(\ub{q}-\ub{p})
                    + H.c.\right\} .  \nonumber
\end{eqnarray}
This is a straightforward generalization of the Hamiltonian
given in \cite{OnePhotonQED}, to include the second PV photon
and the $\xi$ factors.  The vertex functions are the same,
but are repeated here for convenience:
\begin{eqnarray} \label{eq:vertices}
    V^0_{ij\pm}(\ub{p},\ub{q}) &=& \frac{e}{\sqrt{16 \pi^3 }}
                   \frac{ \vec{p}_\perp\cdot\vec{q}_\perp
                      \pm i\vec{p}_\perp\times\vec{q}_\perp
                       + m_i m_j + p^+q^+}{p^+q^+\sqrt{q^+-p^+}} , \\
    V^3_{ij\pm}(\ub{p},\ub{q}) &=& \frac{-e}{\sqrt{16 \pi^3}}
                        \frac{ \vec{p}_\perp\cdot\vec{q}_\perp
                      \pm i\vec{p}_\perp\times\vec{q}_\perp
                       + m_i m_j - p^+q^+ }{p^+q^+\sqrt{q^+-p^+}} , \nonumber\\
    V^1_{ij\pm}(\ub{p},\ub{q}) &=& \frac{e}{\sqrt{16 \pi^3}}
       \frac{ p^+(q^1\pm i q^2)+q^+(p^1\mp ip^2)}{p^+q^+\sqrt{q^+-p^+}} , \nonumber\\
    V^2_{ij\pm}(\ub{p},\ub{q}) &=& \frac{e}{\sqrt{16 \pi^3}}
       \frac{ p^+(q^2\mp i q^1)+q^+(p^2\pm ip^1)}{p^+q^+\sqrt{q^+-p^+}} , \nonumber\\
    U^0_{ij\pm}(\ub{p},\ub{q}) &=& \frac{\mp e}{\sqrt{16 \pi^3}}
       \frac{m_j(p^1\pm ip^2)-m_i(q^1\pm iq^2)}{p^+q^+\sqrt{q^+-p^+}} , \nonumber\\
    U^3_{ij\pm}(\ub{p},\ub{q}) &=& \frac{\pm e}{\sqrt{16 \pi^3}}
       \frac{m_j(p^1\pm ip^2)-m_i(q^1\pm iq^2)}{p^+q^+\sqrt{q^+-p^+}} , \nonumber\\
    U^1_{ij\pm}(\ub{p},\ub{q}) &=& \frac{\pm e}{\sqrt{16 \pi^3}}
                            \frac{m_iq^+-m_jp^+ }{p^+q^+\sqrt{q^+-p^+}} , \nonumber\\
    U^2_{ij\pm}(\ub{p},\ub{q}) &=& \frac{i e}{\sqrt{16 \pi^3}}
                     \frac{m_iq^+-m_jp^+ }{p^+q^+\sqrt{q^+-p^+}} . \nonumber
\end{eqnarray}

\section{One-Photon Truncation}
\label{sec:Truncation}

The dressed-electron problem in QED has been solved analytically for a 
one-photon/one-electron truncation~\cite{OnePhotonQED} in the limit of
an infinite PV electron mass.  For calculations with higher-order
truncation, even for a two-photon truncation, this infinite-mass limit
cannot be taken explicitly.  Therefore, the one-photon truncation
must be studied for finite PV electron masses before proceeding
to higher-order truncations.  The eigenvalue problem is still
analytically soluble; however, there are additional issues to be
addressed in the renormalization, which we discuss in 
Sec.~\ref{sec:Regularization}.

\subsection{Electron Eigenstate}
   
It is convenient to work in a Fock basis where ${\cal P}^+$
and $\vec{\cal P}_\perp$ are diagonal.
We expand the eigenfunction for the dressed-electron state
with total $J_z=\pm \frac12$ in such a Fock basis as
\be \label{eq:FockExpansion}
|\psi^\pm(\ub{P})\rangle=\sum_i z_i b_{i\pm}^\dagger(\ub{P})|0\rangle
  +\sum_{ijs\mu}\int d\ub{k} C_{ijs}^{\mu\pm}(\ub{k})b_{is}^\dagger(\ub{P}-\ub{k})
                                       a_{j\mu}^\dagger(\ub{k})|0\rangle,
\ee
where we keep only the one-electron and one-photon/one-electron Fock sectors
and have chosen the frame where the total transverse momentum is zero.
The amplitudes $z_i$ and wave functions $C_{ijs}^{\mu\pm}$
that define this state must satisfy the coupled 
system of equations that results from the field-theoretic
mass-squared eigenvalue problem (\ref{eq:EigenProb})
and satisfy the normalization condition
\be  \label{eq:norm}
\langle\psi^{\sigma'}(\ub{P}')|\psi^\sigma(\ub{P})\rangle
                  =\delta(\ub{P}'-\ub{P})\delta_{\sigma'\sigma}.
\ee

Careful interpretation of the solution is required to
obtain physically meaningful answers.  In particular,
there needs to be a physical state with positive norm.
We apply the same approach as was used in Yukawa
theory~\cite{YukawaOneBoson}.
A projection onto the physical subspace is accomplished
by expressing Fock states in terms of positively normed
creation operators $a_{0\mu}^\dagger$, $a_{2\mu}^\dagger$, and 
$b_{0s}^\dagger$ and the null combinations 
$a_\mu^\dagger=\sum_i \sqrt{\xi_i}a_{i\mu}^\dagger$ 
and $b_s^\dagger=b_{0s}^\dagger+b_{1s}^\dagger$.
The $b_s^\dagger$ particles are annihilated by the
generalized electromagnetic current $\bar\psi\gamma^\mu\psi$; thus,
$b_s^\dagger$ creates unphysical contributions to be dropped,
and, by analogy, we also drop contributions created by $a_\mu^\dagger$.

The projected dressed-fermion state is
\bea \label{eq:projected}
|\psi_{\rm phys}^\pm(\ub{P})\rangle&=&\sum_i (-1)^i z_i
                                          b_{0\pm}^\dagger(\ub{P})|0\rangle \\
  &&+\sum_{s\mu}\int d\ub{k} \sum_{i=0}^1\sum_{j=0,2}\sqrt{\xi_j}
        \sum_{k=j/2}^{j/2+1} \frac{(-1)^{i+k}}{\sqrt{\xi_k}}
                        C_{iks}^{\mu\pm}(\ub{k})
               b_{0s}^\dagger(\ub{P}-\ub{k})
                                       a_{j\mu}^\dagger(\ub{k})|0\rangle . \nonumber
\eea
This projection is to be used to compute the anomalous moment.  We do not
make the gauge projection $a_{j\mu}^\dagger\rightarrow\tilde a_{j\mu}^\dagger$
defined in (\ref{eq:GaugeProjection}), because gauge invariance has been
broken by both the truncation and the flavor-changing currents.  The remaining
negative norm of $a_{j0}^\dagger$ does not cause difficulties for our
calculations; in particular, the solution has positive norm.

\subsection{Integral Equations}

The amplitudes satisfy coupled equations that come from the 
basic eigenvalue equation $H_{\rm LC}|\psi\rangle=M^2|\psi\rangle$.
These equations are, with $y=k^+/P^+$,
\bea
[M^2-m_i^2]z_i & = & \int P^+ dy d^2k_\perp \sum_{j,l,\mu}\sqrt{\xi_l}(-1)^{j+l}\epsilon^\mu
  P^+\left[V_{ji+}^{\mu*}(\ub{P}-\ub{k},\ub{P})C^{\mu+}_{jl+}(\ub{k}) \right.\\
  &&  \rule{2in}{0mm} \left.+U_{ji+}^{\mu*}(\ub{P}-\ub{k},\ub{P}) C^{\mu+}_{jl-}(\ub{k})\right] , \nonumber
\eea
and
\bea \label{eq:TwoBodyEqns}
\left[M^2 - \frac{m_i^2 + k_\perp^2}{(1-y)} - \frac{\mu_l^2 + k_\perp^2}{y}\right]
  C^{\mu\pm}_{il\pm}(\ub{k}) 
    &=& \sqrt{\xi_l}\sum_j (-1)^j z_j P^+ V_{ij\pm}^\mu(\ub{P}-\ub{k},\ub{P}), \\
\left[M^2 - \frac{m_i^2 + k_\perp^2}{(1-y)} - \frac{\mu_l^2 + k_\perp^2}{y}\right]
C^{\mu\pm}_{il\mp}(\ub{k}) 
    &=& \sqrt{\xi_l}\sum_j (-1)^j z_j P^+ U_{ij\pm}^\mu(\ub{P}-\ub{k},\ub{P}).
\eea

The wave functions $C_{ils}^{\mu\pm}$ are obtained directly~\cite{OnePhotonQED}
\begin{eqnarray} \label{eq:wavefn1}
C^{\mu\pm}_{il\pm}(\ub{k}) &=& \sqrt{\xi_l}
  \frac{\sum_j (-1)^j z_j P^+ V_{ij\pm}^\mu(\ub{P}-\ub{k},\ub{P})}
    {M^2 - \frac{m_i^2 + k_\perp^2}{1-y} - \frac{\mu_l^2 + k_\perp^2}{y}} , \\
\label{eq:wavefn2}
C^{\mu\pm}_{il\mp}(\ub{k}) &=& \sqrt{\xi_l}
\frac{\sum_j (-1)^j z_j P^+ U_{ij\pm}^\mu(\ub{P}-\ub{k},\ub{P})}
     {M^2 - \frac{m_i^2 + k_\perp^2}{1-y} - \frac{\mu_l^2 + k_\perp^2}{y}} .
\end{eqnarray}
These can be eliminated from the first of the coupled equations
to yield
\begin{eqnarray}
(M^2-m_i^2)z_i &=& \int dy\; d^2k_\perp \sum_{\mu,i',j,l}(-1)^{i'+j+l} \xi_l
z_{i'} (P^+)^3 \epsilon^\mu \\
 && \times  \frac{ V^{\mu*}_{ji+}(\ub{P}-\ub{k},\ub{P}) V^\mu_{ji'+}(\ub{P}-\ub{k},\ub{P}) 
             +U^{\mu*}_{ji+}(\ub{P}-\ub{k},\ub{P}) U^\mu_{ji'+}(\ub{P}-\ub{k},\ub{P})}
      {M^2 - \frac{m_j^2 + k_\perp^2}{1-y} - \frac{\mu_l^2 + k_\perp^2}{y}} ,
      \nonumber
\end{eqnarray}
which, on use of the definitions (\ref{eq:vertices}) of the vertex functions,
can be written more usefully as
\be \label{eq:FeynEigen}
(M^2-m_i^2)z_i =
      2e^2\sum_{i'} (-1)^{i'}z_{i'}\left[\bar{J}+m_im_{i'} \bar{I}_0
  -2(m_i+m_{i'}) \bar{I}_1 \right],
\ee
with
\begin{eqnarray} \label{eq:In}
\bar{I}_n(M^2)&=&\int\frac{dy dk_\perp^2}{16\pi^2}
   \sum_{jl}\frac{(-1)^{j+l}\xi_l}{M^2-\frac{m_j^2+k_\perp^2}{1-y}
                                   -\frac{\mu_l^2+k_\perp^2}{y}}
   \frac{m_j^n}{y(1-y)^n}\,, \\
\bar{J}(M^2)&=&\int\frac{dy dk_\perp^2}{16\pi^2}  \label{eq:J}
   \sum_{jl}\frac{(-1)^{j+l}\xi_l}{M^2-\frac{m_j^2+k_\perp^2}{1-y}
                                   -\frac{\mu_l^2+k_\perp^2}{y}}
   \frac{m_j^2+k_\perp^2}{y(1-y)^2} .
\end{eqnarray}
The form of (\ref{eq:FeynEigen}) matches that of the equivalent eigenvalue 
problem in Yukawa theory~\cite{YukawaOneBoson}, with the replacements 
$g^2\rightarrow 2e^2$, $\mu_0 I_1\rightarrow -2\bar{I}_1$, and
$\mu_0^2 J\rightarrow \bar{J}$.  

The integrals $\bar{I}_0$ and $\bar{J}$ satisfy an identity,
$\bar{J}=M^2 \bar{I}_0$.  This was stated in \cite{OnePhotonQED}
without a proof being given.  A new, simple proof can be found
in Appendix~\ref{sec:Identity} of this paper.  With use of this identity,
the eigenvalue problem reduces to the simpler form
\be \label{eq:FeynEigen2}
(M^2-m_i^2)z_i =
      2e^2\sum_{i'} (-1)^{i'}z_{i'}\left[(M^2+m_im_{i'}) \bar{I}_0
  -2(m_i+m_{i'}) \bar{I}_1 \right].
\ee

\subsection{Solution of the Eigenvalue Problem}
 
The solution to the eigenvalue problem is~\cite{OnePhotonQED}
\begin{equation} \label{eq:OneBosonEigenvalueProb}
\alpha_\pm=\frac{(M\pm m_0)(M\pm m_1)}{8\pi (m_1-m_0)(2 \bar{I}_1\pm M\bar{I}_0)} , \;\;
z_1=\frac{M \pm m_0}{M \pm m_1}z_0 ,
\end{equation}
with $z_0$ determined by normalization.  
The simplicity of this result is due in part to the algebraic
simplification of (\ref{eq:FeynEigen}) that comes from the identity
$\bar J=M^2\bar I_0$.  

The value of $m_0$ is
determined by requiring $\alpha_\pm$ to be equal to the
physical value of $\alpha$.  For small values of the PV masses
there may be no such solution; however, for reasonable values
we do find at least one solution for each branch.

\begin{figure}[bhtp]
\centerline{\includegraphics[width=12cm]{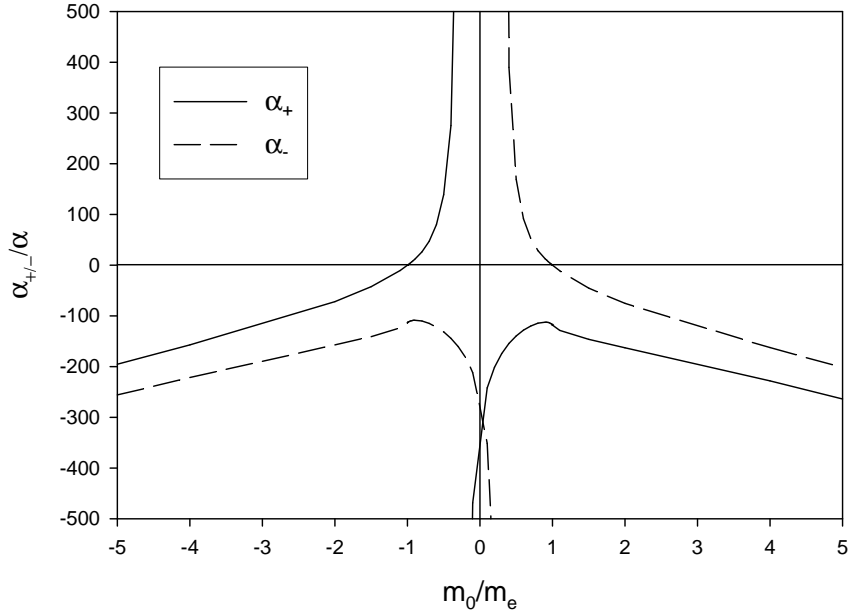}}
\caption{\label{fig:alphapm} The two solutions of the one-photon
eigenvalue problem, for PV masses $m_1=1000 m_e$, $\mu_1=10 m_e$,
and $\mu_2=\infty$. 
The horizontal line shows where $\alpha_\pm=\alpha$. 
The $\alpha_-$ branch corresponds to the 
physical choice, but with $m_0$ less than $m_e$.}
\end{figure}

The plot in Fig.~\ref{fig:alphapm} shows $\alpha_\pm/\alpha$
as functions of $m_0$.  
The $\alpha_-$ branch is the
physical choice, because the no-interaction limit ($\alpha_-=0$)
corresponds to the
bare mass $m_0$ becoming equal to the physical electron
mass, $M=m_e$.  

If the PV electron has a sufficiently large mass,
the value of $m_0$ that yields $\alpha_-=\alpha$ is
less than $m_e$.  In this case, the integrals $\bar I_n$
and $\bar J$ contain poles for $j=l=0$ and are defined
by a principal-value prescription~\cite{OnePhotonQED}.
The presence of the poles can then admit an additional
delta-function term to the two-body wave function:
\be
C_{00s}^{\mu\sigma}(\ub{k}) \rightarrow
 C_{00s}^{\mu\sigma}(\ub{k})+c_s^{\mu\sigma}\delta(\ub{k}-\ub{k}_0),
\ee
where $\ub{k}_0$ is such that 
$M^2=\frac{m_0^2 + k_{0\perp}^2}{1-y_0} + \frac{\mu_0^2 + k_{0\perp}^2}{y_0}$.
This remains a solution to (\ref{eq:TwoBodyEqns}), but there
will be additional terms in (\ref{eq:FeynEigen}) proportional
to $c_s^{\mu\sigma}$.
We do not explore this possibility, because we have
found that when we include self-energy corrections 
from a two-photon truncation, the poles in these wave
functions disappear.

\subsection{Normalization}  \label{sec:normalization}

The normalization of the wave functions is determined
by the condition in Eq.~(\ref{eq:norm}).  In the case
of the present truncation, this reduces to
\begin{equation}
1=|\sum_i(-1)^iz_i|^2
+\sum_{s\mu}\int d\ub{k}\epsilon^\mu \sum_{j=0,2}\xi_j
   \left|\sum_{i=0}^1\sum_{k=j/2}^{j/2+1}
    \frac{(-1)^{i+k}}{\sqrt{\xi_k}}C_{iks}^{\mu+}(\ub{k})\right|^2.
\end{equation}
For the given wave functions, after some tedious calculations, this becomes
\bea
\frac{1}{z_0^2}&=&(1-\zeta_1)^2 \\
&& +\frac{\alpha}{2\pi}\int y dy dk_\perp^2\sum_{l,l'}(-1)^{l+l'}\zeta_l \zeta_{l'}
  \sum_{i'i}(-1)^{i'+i}\sum_{j=0,2}\xi_j\sum_{k'=j/2}^{j/2+1}\sum_{k=j/2}^{j/2+1}
     (-1)^{k'+k} \nonumber \\
 && \times
    \frac{m_{i'} m_i-(m_i+m_{i'})(m_l+m_{l'})(1-y)+m_lm_{l'}(1-y)^2+k_\perp^2}
  {[ym_{i'}^2+(1-y)\mu_{k'}^2+k_\perp^2-m_e^2y(1-y)]
                [ym_i^2+(1-y)\mu_k^2+k_\perp^2-m_e^2y(1-y)]} ,
  \nonumber
\eea
where $\zeta_l=z_l/z_0$.  

For terms with $i=k=0$ or $i'=k'=0$, there are
simple poles defined by a principal-value prescription.  For the terms
where all four of these indices are zero, there is a double pole,
defined by the prescription~\cite{OnePhotonQED}
\begin{equation} \label{eq:prescrip}
\int dx \frac{f(x)}{(x-a)^2}\equiv  \lim_{\eta\rightarrow 0} \frac{1}{2\eta}
  \left[\mathcal{P}\int dx\frac{f(x)}{x-a-\eta}
          -\mathcal{P}\int dx \frac{f(x)}{x-a+\eta}\right].
\end{equation}
One could instead compute the norm by taking the zero-momentum limit
of the Dirac form factor, $F_1$; however, this would correspond to a 
more complicated point splitting.  Our prescription splits only with
respect to the magnitude of the momentum, rather than the magnitude
and angle.

\section{Regularization and Renormalization}
\label{sec:Regularization}

To evaluate the usefulness of the chosen regularization,
we consider three aspects.  One is to compare the result
for the one-loop electron self-energy with the standard result from
covariant Feynman theory; we do this indirectly, by first 
comparing with the infinite-momentum-frame result of Brodsky,
Roskies, and Suaya~\cite{BRS}, which they show to be consistent
with Feynman theory.  The second is to check the massless
chiral limit, where we find a specific constraint on the
PV photon masses and couplings.  The third is to consider
the renormalization of the external coupling to the charge.
We exclude fermion-antifermion states, and, therefore,
there is no vacuum polarization.  Thus, if the vertex and
wave function renormalizations cancel, there will be no
renormalization of the external coupling.  This is what
we find, but only for the plus component of the current.

In our formulation, the perturbative one-loop electron self-energy
can be read from Eq.~(\ref{eq:FeynEigen}) for $i=0$, with
$z_1=0$, $M^2=m_0^2+\delta m^2$ on the left, and $M^2=m_0^2$
on the right.  This yields
\be
\delta m^2=2e^2\left[\bar J(m_0^2)+m_0^2\bar I_0(m_0^2)-4m_0\bar I_1(m_0^2)\right].
\ee
When $\delta m =\delta m^2/2m_0$ is written explicitly in
terms of $\alpha=e^2/4\pi$ and the integrals (\ref{eq:In})
and (\ref{eq:J}), we have
\be \label{eq:deltam0}
\delta m=\frac{\alpha}{4\pi}\sum_{jl}(-1)^{j+l}\frac{\xi_l}{m_0}
   \int\frac{dy}{y}\frac{d^2k_\perp}{\pi} \frac{m_0^2-\frac{4m_0m_j}{1-y}+\frac{m_j^2+k_\perp^2}{(1-y)^2}}
      {m_0^2-\frac{m_0^2+k_\perp^2}{1-y}-\frac{\mu_l^2+k_\perp^2}{y}}.
\ee
To compare with \cite{BRS}, where the self-energy is regulated with
only one PV photon, we restrict the sum over $l$ to two terms, $l=0$
and $l=1$.  In this case, the $j=0$ term matches the form of $\delta m_a$ 
in Eq.~(3.40) of \cite{BRS},\footnote{There is some discussion of these points
in \protect\cite{OnePhotonQED}, though for a different regularization.
There is, however, a sign error in the corresponding equation
of \protect\cite{OnePhotonQED}, Eq. (39); the polynomial in the numerator
should be $(1-4x+x^2)$.  Also, the right-hand sides of both (39) and
(40) should be divided by $m$, and the left-hand sides should read
$\delta m_a$ and $\delta m_b$, respectively.}
which we quote here
\be
\delta m_a=\frac{e^2}{16\pi^2 m_0} \int d^2k_\perp \int\frac{dx}{1-x}
\left[ \frac{m_0^2(2-2x-x^2)-k_\perp^2}{\lambda^2(1-x)+k_\perp^2+m_0^2 x^2}
      -\frac{m_0^2(2-2x-x^2)-k_\perp^2}{\Lambda^2(1-x)+k_\perp^2+m_0^2 x^2}\right].
\ee
The $j=0$ term of (\ref{eq:deltam0}) takes this form after setting
$y=1-x$, $\mu_0=\lambda$, and $\mu_1=\Lambda$
and making some algebraic rearrangements.  Also, the $j=1$ term
reduces to $\delta m_b$ in Eq.~(3.41) of \cite{BRS} in the limit 
$m_1\rightarrow\infty$.  In general, it is in this limit that the
instantaneous fermion contributions return to the theory, and the
source of $\delta m_b$ is just this type of graph.  Here we do not
take this limit, and the $j=1$ term remains as written and yields
a different form for $\delta m_b$.  However, if Brodsky {\em et al}.
had used our regularization, they would also obtain this different form.
Thus, our regularization produces a one-loop self-energy correction
which is consistent with \cite{BRS} when the same regularization
is used, namely one PV electron and two PV photons, since the
subtractions of contributions from the PV particles have exactly
the same forms.  This, in turn, is consistent with the Feynman result.

Although consistent with the standard result when regulated in
the same way, the one-loop self-energy in this regularization,
whether by covariant methods, by the infinite-momentum frame approach, or
by light-cone quantization, does not automatically have the
correct massless limit of zero, and chiral symmetry is broken.
Consider the mass shift $\delta m$ in terms of the integrals 
defined in (\ref{eq:In}) and (\ref{eq:J}).  From (\ref{eq:deltam0})
we have
\be
\delta m=16\pi^2\frac{\alpha}{2\pi}\left[m_0 \bar{I}_0(m_0^2)-2 \bar{I}_1(m_0^2)\right].
\ee
In the chiral limit, $m_0\rightarrow0$, we obtain
\be
\delta m=-32\pi^2\frac{\alpha}{2\pi}\bar{I}_1(0),
\ee
with
\be
\bar{I}_1(0)=\frac{m_1}{16\pi^2}\sum_l (-1)^l\xi_l\int dy d^2k_\perp
       \frac{1}{k_\perp^2+m_1^2 y+\mu_l^2(1-y)}.
\ee
The integrals in $\bar{I}_1(0)$ can be easily done, to find
\be
\delta m=-\frac{\alpha}{\pi}m_1\sum_l(-1)^l\xi_l
   \frac{\mu_l^2/m_1^2}{1-\mu_l^2/m_1^2}\ln(\mu_l^2/m_1^2) .
\ee
Clearly, this is zero only if $m_1$ is infinite or the $\xi_l$
and masses $\mu_l$ satisfy the constraint
\be
\sum_l(-1)^l\xi_l
   \frac{\mu_l^2/m_1^2}{1-\mu_l^2/m_1^2}\ln(\mu_l^2/m_1^2)=0 .
\ee
This cannot be satisfied without the introduction of a second PV photon.

When the PV electron mass is sufficiently large, the
chiral-limit constraint can be approximated by
\be \label{eq:chiralconstraint}
\sum_l(-1)^l\xi_l\mu_l^2\ln(\mu_l/m_1)=0. 
\ee
The solution to the set of constraints, Eq.~(\ref{eq:nullconstraint})
and (\ref{eq:chiralconstraint}) along with $\xi_0=1$ and $\mu_0=0$, is then
\be
\xi_1=1+\xi_2 \;\;  \mbox{and} \;\; 
\xi_2=\frac{\mu_1^2\ln(\mu_1/m_1)}
     {\mu_2^2\ln(\mu_2/m_1)-\mu_1^2\ln(\mu_1/m_1)}.
\ee
Without loss of generality, we require $\mu_2>\mu_1$, so that $\xi_2$
is positive.

In covariant perturbation theory, it is a consequence of the Ward identity
that, order by order, the wave function renormalization constant $Z_2$
is equal to the vertex renormalization $Z_1$.  As discussed in \cite{BRS},
this equality holds true more generally for nonperturbative bound-state 
calculations.  However, a Fock-space truncation can have the effect of
destroying covariance of the electromagnetic current, so that some
components of the current require renormalization despite the
absence of vacuum polarization.  In the particular case here, only 
couplings to the plus component are not renormalized.  The lack of
fermion-antifermion vertices destroys covariance.

To see that $Z_1=Z_2$ holds for the plus component, define a bare
state $|\psi_{\rm bare}\rangle$ of the electron as a Fock-state expansion
in which the one-electron state has amplitude 1.  It is then related
to the physical electron state by
\be 
|\psi_{\rm phys}\rangle=\sqrt{Z_2}|\psi_{\rm bare}\rangle.
\ee
The normalization of the physical state 
$\langle\psi_{\rm phys}(p')|\psi_{\rm phys}(p)\rangle =\delta(\ub{p}'-\ub{p})$
implies
\be
\langle\psi_{\rm bare}(p')|\psi_{\rm bare}(p)\rangle =Z_2^{-1}\delta(\ub{p}'-\ub{p}).
\ee
Matrix elements of the current $J^\mu$ define $Z_1$ by 
\be
\langle\psi_{\rm bare}|J^\mu(0)|\psi_{\rm bare}\rangle=Z_1^{-1}\bar u(p)\gamma^\mu u(p).
\ee
For the plus component, this matrix element can also be calculated 
as~\cite{BrodskyDrell}
\be
\langle\psi_{\rm bare}(p')|J^+(0)|\psi_{\rm bare}(p)\rangle=2p^+F_{1{\rm bare}}(-(p'-p)^2).
\ee
Because~\cite{LepageBrodsky} $\bar u(p)\gamma^+ u(p)=2p^+$ and $F_{1{\rm bare}}(0)=Z_2^{-1}$,
we find that $\langle\psi_{\rm bare}|J^+(0)|\psi_{\rm bare}\rangle$ is equal to
both $2p^+Z_2^{-1}$ and $2p^+Z_1^{-1}$, and therefore we have $Z_1=Z_2$.

The calculation of the anomalous moment, discussed in the next section, can then
proceed.  It is based on matrix elements of the plus component and thus does not
require additional renormalization.

\section{Anomalous Magnetic Moment}
\label{sec:AnomalousMoment} 

We start from the Brodsky--Drell formula
for the anomalous moment derived in \cite{BrodskyDrell}
from the spin-flip matrix element of the electromagnetic current.
In the one-photon truncation their formula reduces to
\bea
a_e&=&m_e\sum_{s\mu}\int d\ub{k}\epsilon^\mu \sum_{j=0,2}\xi_j
  \left(\sum_{i'=0}^1\sum_{k'=j/2}^{j/2+1}
    \frac{(-1)^{i'+k'}}{\sqrt{\xi_{k'}}}C_{i'k's}^{\mu+}(\ub{k})\right)^* \\
  && \times y\left(\frac{\partial}{\partial k_x}+i\frac{\partial}{\partial k_y}\right)
  \left(\sum_{i=0}^1\sum_{k=j/2}^{j/2+1}
    \frac{(-1)^{i+k}}{\sqrt{\xi_k}}C_{iks}^{\mu-}(\ub{k})\right). \nonumber
\eea
The presence of the derivative of the wave function 
(see Eqs.~(\ref{eq:wavefn1}) and (\ref{eq:wavefn2}))
implies that we may face a triple
pole; however, these terms cancel, and the expression for the anomalous
moment simplifies to
\bea
a_e&=&\frac{\alpha}{\pi}m_e\int y^2 (1-y) dy dk_\perp^2
\sum_{l,l'}(-1)^{l+l'}z_l z_{l'}m_l\sum_{j=0,2}\xi_j \\
  &&  \times\left(\sum_{i=0}^1\sum_{k=j/2}^{j/2+1} \frac{(-1)^{i+k}}{ym_i^2+(1-y)\mu_k^2+k_\perp^2-m_e^2y(1-y)}\right)^2 . \nonumber
\eea
The double pole is handled in the same way as for the normalization
integrals, discussed in Sec.~\ref{sec:normalization}.
The integrals can be done analytically.

In the limit where the PV electron mass $m_1$ is infinite,
the bare-electron
amplitude ratio $z_1/z_0$ is zero but the limit of the
product $m_1z_1/z_0$ is $m_0-m_e$.  Thus,
the limit of the expression for the anomalous moment is
\begin{equation}
a_e=\frac{\alpha}{\pi}m_e^2z_0^2\int y^2 (1-y) dy dk_\perp^2
    \sum_{j=0,2}\xi_j\left(\sum_{k=j/2}^{j/2+1} \frac{(-1)^k}{ym_0^2+(1-y)\mu_k^2+k_\perp^2-m_e^2y(1-y)}\right)^2 .
\end{equation}
This differs slightly from the expression given in Eq.~(70) of \cite{OnePhotonQED},
where only one PV photon was included,
the projection onto physical states was not taken,
and $m_1z_1$ was assumed to be zero; however, the difference
in values is negligible when $\mu_1$ and $\mu_2$ are sufficiently large.

\begin{figure}[tbhp]
\centerline{\includegraphics[width=15cm]{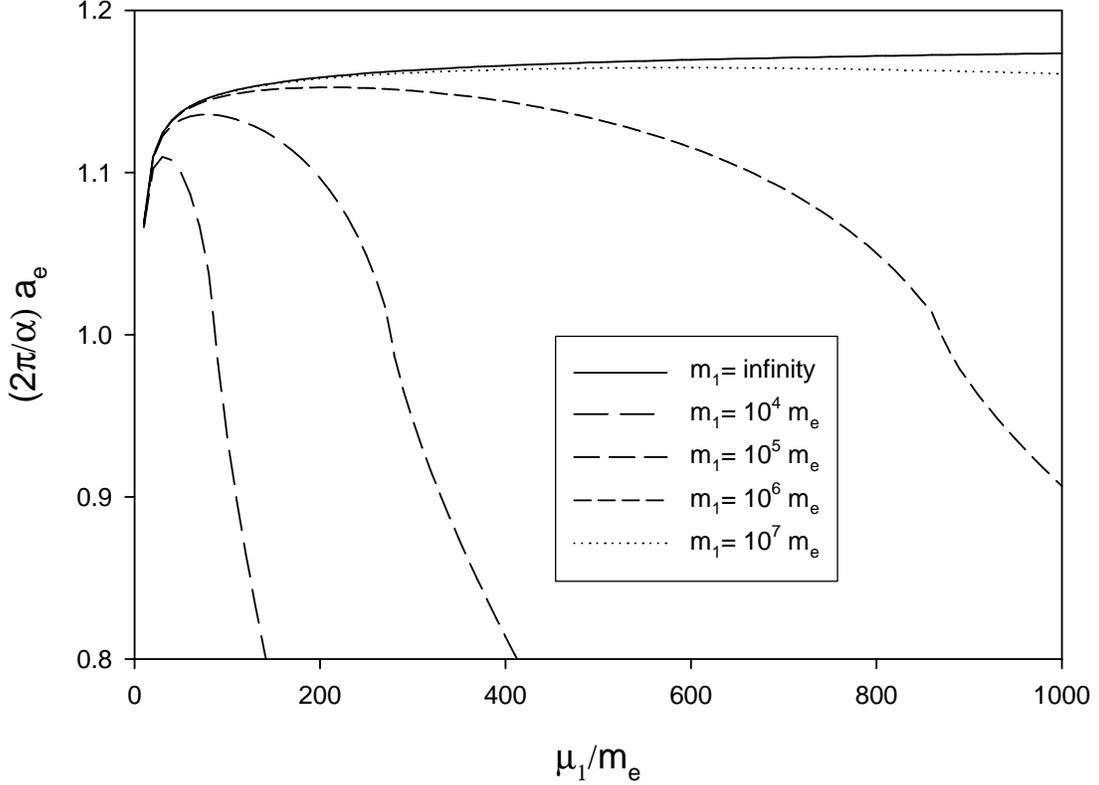}}
\caption{\label{fig:aeOnePhoton} The anomalous moment of the electron in
units of the Schwinger term (${\alpha/2\pi}$) plotted versus
the PV photon mass, $\mu_1$, for a few values of the PV electron mass,
$m_1$. The second PV photon is absent, and the chiral symmetry of the
massless limit is broken by the remaining regularization.}
\end{figure}

If the second PV photon is not included,
the results for the anomalous moment have a very strong dependence on the
PV masses $\mu_1$ and $m_1$, as shown in Fig.~\ref{fig:aeOnePhoton}.
A slowly varying behavior with respect to the PV photon mass $\mu_1$
is obtained only if the PV electron mass $m_1$ is (nearly) infinite.

The strong variation with $\mu_1$ when $m_1$ is finite is a 
consequence of broken chiral symmetry.  This can be seen as
follows.  The anomalous moment is very sensitive to
the masses of the constituents~\cite{hb}, and the mass $m_0$ of the
electron constituent is determined by the eigenvalue solution
(\ref{eq:OneBosonEigenvalueProb}), which contains the integral $\bar{I}_1$.  
Relative to the integral's value $\bar{I}_{1\infty}$ at infinite $m_1$,
we have
\be
\bar I_1=\bar I_{1\infty}+\int\frac{dy dk_\perp^2}{16\pi^2}
  \sum_j \frac{(-1)^j \xi_j m_1}{k_\perp^2+ym_1^2+(1-y)\mu_j^2-m_e^2 y(1-y)}.
\ee
When $m_e$ is neglected compared to $m_1$, the second term becomes
the chiral limit of $\bar{I}_1$, and this introduces a correction
to the bare-electron mass of the form 
$\frac{\mu_1^2 \ln (\mu_1/m_1)}{8\pi^2 m_e m_1(1-\mu^2/m_1^2)}$.
This correction is removed when the second PV photon is included,
because the chiral limit of $\bar{I}_1$ is then zero, but when
the correction is not removed, it injects a very strong dependence
on $\mu_1$ and $m_1$ into the behavior of the bare mass $m_0$ and
thus into the behavior of the anomalous moment.

From Fig.~\ref{fig:aeOnePhoton}, we see that, without the second
PV photon, the PV electron mass needs to be on the order of 
$10^7\,m_e$ before results for
the one-photon truncation approach the infinite-mass limit.
Thus, we estimate that the PV electron mass must be at least this large
for a calculation with a two-photon Fock-space truncation, if only 
one PV photon is included.  Unfortunately, such large mass values 
make numerical calculations difficult, because of contributions to 
integrals at momentum fractions of order $(m_e/m_1)^2\simeq 10^{-14}$, 
which are then subject to large round-off errors.  Therefore,
a practical two-photon calculation will require the second PV photon.

\begin{figure}[tbhp]
\centerline{\includegraphics[width=15cm]{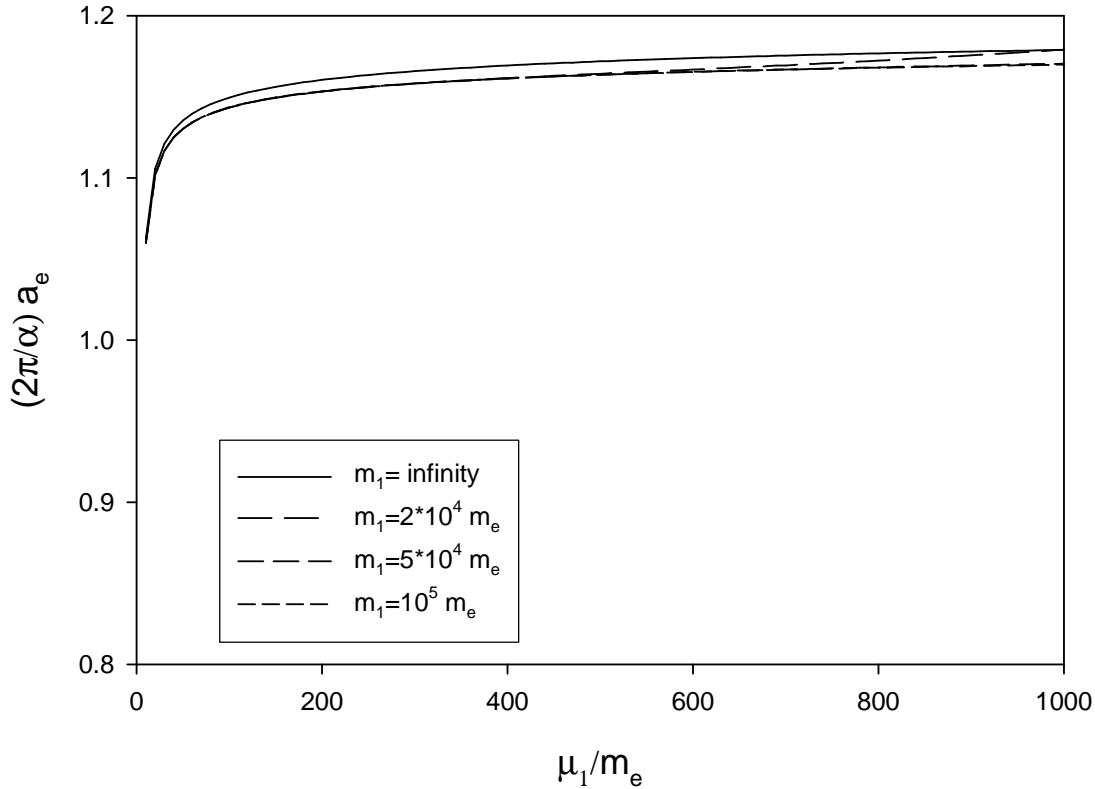}}
\caption{\label{fig:ae2OnePhoton} Same as Fig.~\ref{fig:aeOnePhoton},
but with the second PV photon included, with a mass $\mu_2=\sqrt{2}\mu_1$,
and the chiral symmetry is restored.
The mass ratio is held fixed as $\mu_1$ and $\mu_2$ are varied.}
\end{figure}

We now repeat the calculation of the anomalous moment in the
one-photon truncation with the second PV photon included.  The
result is given in Fig.~\ref{fig:ae2OnePhoton} for PV masses
related by $\mu_2=\sqrt{2}\mu_1$.
Clearly, the dependence on the PV masses is greatly reduced.  The
value obtained for the anomalous moment differs from the leading-order
Schwinger result~\cite{Schwinger}, and thus from the physical value, by 17\%.  
Agreement at this level of accuracy is to be expected; the leading
divergence in the normalization will not be cancelled until the
truncation is relaxed to include two-photon states.

\section{Summary}
\label{sec:Summary}

We have developed an improved Pauli--Villars regularization
of light-front QED by establishing the correct chiral limit.
Chiral symmetry is restored by the introduction of an additional
Pauli--Villars photon.  An application to the calculation of
the anomalous magnetic moment of the electron shows much
less dependence on the regulator masses, as can be seen in
comparing results without and with the second PV field in
Figs.~\ref{fig:aeOnePhoton} and \ref{fig:ae2OnePhoton},
respectively.

The chiral condition on the coupling and mass of the second
PV field is given in Eq.~(\ref{eq:chiralconstraint}).  It is
of the same form as the constraint obtained earlier for Yukawa
theory with three PV bosons~\cite{ChiralSymYukawa,bhm1}.
Such a constraint should also be considered for the regularization
of Yukawa theory with one PV boson and one PV fermion.  This was 
not done in \cite{YukawaOneBoson} or \cite{YukawaTwoBoson}; however, 
there none of the bosons is massless, and the anomalous moment is 
much less sensitive to constituent masses.

For less severe truncations, where more photons are allowed in
Fock states, the number of PV photon flavors does not need to
increase~\cite{Paston}.  However, the chiral constraint will
be more complicated.  Fortunately, the corrections
will be higher order in $\alpha$, and therefore 
should be small enough to be neglected.

Thus, we have a regularization scheme that can properly handle
the one-photon truncation at finite PV electron mass and can
be readily extended to higher truncations.  The result for
the anomalous moment in the one-photon truncation does differ
by 17\% from the experimental result, but this discrepancy is
expected to be much reduced in the two-photon truncation
which includes self-energy effects for the constituent electron.

\acknowledgments
This work was supported by the Department of Energy
through Contracts No.\ DE-FG03-95ER40908 (S.S.C.) and
No.\ DE-FG02-98ER41087 (J.R.H.).  The paper is dedicated
to the memory of Gary McCartor, who originated this
particular approach to Pauli--Villars regularization
and continued to work on its applications until his
untimely passing.

\appendix

\section{Gauge Condition}
\label{sec:GaugeCondition}

The gauge condition $\partial^\mu A_{i\mu}=0$ can be implemented
as a projection for the positive frequency part~\cite{GuptaBleuler,GaugeCondition},
with physical states $|\psi\rangle$ restricted by 
\be
\partial^\mu A_{i\mu}^{(+)}|\psi\rangle
=\frac{1}{\sqrt{16\pi^3}}\int \frac{d\ub{k}}{\sqrt{k^+}}
      k^\mu a_{i\mu}(\ub{k})e^{-i\ub{k}\cdot\ub{x}}|\psi\rangle=0 .
\ee
This restricts Fock-state expansions to physical polarizations in the
following way~\cite{GaugeCondition}.  Let $e_\mu^{(\lambda)}(\ub{k})$, with $\lambda=0,1,2,3$,
be polarization vectors for a photon with four-momentum $k$, with the properties
\be \label{eq:PolarizationConditions}
e^{(\lambda)\mu}e_\mu^{(\lambda')}=-\epsilon^\lambda\delta_{\lambda\lambda'},
\ee
and
\be
k^\mu e_\mu^{(\lambda)}=0, \;\; n^\mu e_\mu^{(\lambda)}=0, \;\; \lambda=1,2.
\ee
Here $\epsilon^\mu=(-1,1,1,1)$ is the metric signature, as in Sec.~II,
and $n$ is the timelike four-vector that reduces to $(1,0,0,0)$ in the frame
where $\vec k_\perp=0$.  We express the annihilation operator $a_{i\mu}$
in terms of these polarizations as
\be
a_{i\mu}=\sum_\lambda e_\mu^{(\lambda)} a_i^{(\lambda)}.
\ee
The polarizations $\lambda=1,2$ are the physical transverse polarizations.
The scalar and longitudinal polarizations may be chosen to be~\cite{GaugeCondition}
\be \label{eq:UnphysicalPolarizations}
e^{(0)}=n \;\; \mbox{and} \;\; e^{(3)}(\ub{k})=\frac{k-(k\cdot n)n}{k\cdot n} ,
\ee
which satisfy the conditions (\ref{eq:PolarizationConditions}).  From these
choices for $e^{(0)}$ and $e^{(3)}$,
we have
\be
k^\mu e_\mu^{(0)}=k\cdot n , \;\; k^\mu e_\mu^{(3)}=-k\cdot n ,
\ee
and
\be
k^\mu a_{i\mu}=k\cdot n(a_i^{(0)}-a_i^{(3)}) .
\ee
Given this last result, it is convenient to 
define the linear combinations 
\be
a_i^{(\pm)}=(a_i^{(0)}\pm a_i^{(3)})/\sqrt{2} .
\ee
They are both null and satisfy the commutation relations
\be  \label{eq:pmCommutators}
[a_i^{(\pm)}(\ub{k}),a_i^{(\pm)\dagger}(\ub{k}')]=0, \;\;
[a_i^{(\pm)}(\ub{k}),a_i^{(\mp)\dagger}(\ub{k}')]=-(-1)^i\delta(\ub{k}-\ub{k}') .
\ee
The restriction on physical states then reduces to
\be \label{eq:GaugeRestriction}
a_i^{(-)}|\psi\rangle=0 .
\ee
Because $a_i^{(-)}$ commutes with all but $a_i^{(+)\dagger}$, the
restriction (\ref{eq:GaugeRestriction}) can be satisfied by removing 
from $|\psi\rangle$ all
terms that contain $a_i^{(+)\dagger}$.  This is accomplished by
replacing all photon creation operators $a_{i\mu}^\dagger$ with the
projected operator
\be \label{eq:GaugeProjection}
\tilde a_{i\mu}^\dagger=
    \frac{1}{\sqrt{2}}(e_\mu^{(0)}-e_\mu^{(3)})a_i^{(-)\dagger}
  +\sum_{\lambda=1}^2e_\mu^{(\lambda)}a_i^{(\lambda)\dagger} .
\ee
Since the $a_i^{(-)\dagger}$ are null, only the physical polarizations
contribute to expectation values of physical quantities.

The presence of photons created by $a_i^{(-)\dagger}$ corresponds to the
residual gauge transformations~\cite{GaugeCondition} 
that satisfy $\partial^\mu A_{i\mu}=0$,
where $A_{i\mu}\rightarrow A_{i\mu}+\partial_\mu\Lambda_i$ with $\Box\Lambda_i=0$.
To see this, consider the expectation value~\cite{GaugeCondition} 
$\langle\psi|A_{i\mu}|\psi\rangle$, with $|\psi\rangle$ written as
\be
|\psi\rangle=C_0|0\rangle+\int d\ub{q} C_1(\ub{q})a_i^{(-)\dagger}(\ub{q})|0\rangle+\cdots ,
\ee
and transverse polarizations absent.
In the expectation value only the $a_i^{(+)}$ and $a_i^{(+)\dagger}$
terms of $A_{i\mu}$ can contribute, as follows from the commutators
in Eq.~(\ref{eq:pmCommutators}), and these terms give
\be
\langle\psi|A_{i\mu}|\psi\rangle=
  -(-1)^i C_0^*\int\frac{d\ub{q}}{\sqrt{16\pi^3q^+}}
        C_1(\ub{q})e^{-iq\cdot x}
        \frac{1}{\sqrt{2}}(e_\mu^{(0)}(\ub{q})+e_\mu^{(3)}(\ub{q}))+\mbox{c.c.}
\ee
From (\ref{eq:UnphysicalPolarizations}) we have
$e_\mu^{(0)}(\ub{q})+e_\mu^{(3)}(\ub{q})=2q_\mu/q^+$.
The factor $q_\mu$ can be replaced by a partial derivative, leaving
\be
\langle\psi|A_{i\mu}|\psi\rangle=\partial_\mu \Lambda_i(x),
\ee
with
\be
\Lambda_i(x)=-i(-1)^i C_0^*\int\frac{d\ub{q}}{(2\pi q^+)^{3/2}}
                      C_1(\ub{q})e^{-iq\cdot x}+\mbox{c.c.}
\ee
Since $q$ is null, $\Box\Lambda_i=0$.
Thus, the contribution from the unphysical polarizations is a pure gauge
term consistent with the residual gauge symmetry.  A choice of wave
function for the minus polarization corresponds to a choice for the
residual gauge.

\section{Proof of $\bar{J}=M^2 \bar{I}_0$}
\label{sec:Identity}

Here we give a proof of the identity $\bar{J}=M^2 \bar{I}_0$
for the integrals $\bar{I}_0$ and $\bar{J}$ defined in
(\ref{eq:In}) and (\ref{eq:J}), respectively. 
It involves an interesting coordinate transformation
that might have broader application.

We write the integrals in terms of
their individual Fock-sector contributions as
\bea
\bar I_0&=&-\frac{1}{16\pi^2}\sum_{jl}(-1)^{j+l}\xi_l I_{0jl} ,\\
\bar J&=&-\frac{1}{16\pi^2}\sum_{jl}(-1)^{j+l}\xi_l J_{jl} , \nonumber
\eea
with
\bea
I_{0jl}&\equiv &\int \frac{dy dk_\perp^2}{y}
   \frac{1}{\frac{m_j^2+k_\perp^2}{1-y}+\frac{\mu_l^2+k_\perp^2}{y}-M^2}, \\
J_{jl}&\equiv &\int \frac{dy dk_\perp^2}{y(1-y)^2}
   \frac{m_j^2+k_\perp^2}{\frac{m_j^2+k_\perp^2}{1-y}+\frac{\mu_l^2+k_\perp^2}{y}-M^2} .
\nonumber
\eea
For the $J$ integrals, we replace $y$ with a new variable $x$ defined by
\be
x=(1-y)\frac{\mu_l^2+k_\perp^2}{m_j^2 y+\mu_l^2 (1-y)+k_\perp^2} .
\ee
It also ranges between 0 and 1, though in the reverse order relative to $y$, and has
the remarkable property that
\be
\frac{m_j^2+k_\perp^2}{1-y}+\frac{\mu_l^2+k_\perp^2}{y}
  =\frac{m_j^2+k_\perp^2}{1-x}+\frac{\mu_l^2+k_\perp^2}{x},
\ee
even though $x$ and $y$ are clearly not equal and are not even
linearly related.  

With this change of variable, the $J$ integrals become
\be
J_{jl}=\int \frac{dx dk_\perp^2}{x}
   \frac{m_j^2 x+\mu_l^2 (1-x)+k_\perp^2}{x(1-x)}
   \frac{1}{\frac{m_j^2+k_\perp^2}{1-x}+\frac{\mu_l^2+k_\perp^2}{x}-M^2}.
\ee
The middle factor can be written as
\be
\frac{m_j^2 x+\mu_l^2 (1-x)+k_\perp^2}{x(1-x)}
  =\frac{m_j^2+k_\perp^2}{1-x}+\frac{\mu_l^2+k_\perp^2}{x}-M^2+M^2,
\ee
so that we obtain
\be
J_{jl}= \int \frac{dx dk_\perp^2}{x}
   +\int \frac{dx dk_\perp^2}{x}
   \frac{M^2}{\frac{m_j^2+k_\perp^2}{1-x}+\frac{\mu_l^2+k_\perp^2}{x}-M^2}.
\ee

This last result shows that $J_{jl}$ is just $M^2 I_{0jl}$ plus 
an (infinite) constant.  Since the constant cancels in the sum over 
PV particles $\sum_{jl}(-1)^{j+l}\xi_l$, we have the desired identity
of $\bar J=M^2\bar I_0$.


\end{document}